\documentclass[10pt,a4paper,onecolumn]{article}
\usepackage{marginnote}
\usepackage{graphicx}
\usepackage{xcolor}
\usepackage{authblk,etoolbox}
\usepackage{titlesec}
\usepackage{calc}
\usepackage{tikz}
\usepackage{hyperref}
\hypersetup{colorlinks,breaklinks=true,
            urlcolor=[rgb]{0.0, 0.5, 1.0},
            linkcolor=[rgb]{0.0, 0.5, 1.0}}
\usepackage{caption}
\usepackage{tcolorbox}
\usepackage{amssymb,amsmath}
\usepackage{ifxetex,ifluatex}
\usepackage{seqsplit}
\usepackage{xstring}

\usepackage{float}
\let\origfigure\figure
\let\endorigfigure\endfigure
\renewenvironment{figure}[1][2] {
    \expandafter\origfigure\expandafter[H]
} {
    \endorigfigure
}

\usepackage{fixltx2e} 
\usepackage[
  backend=biber,
]{biblatex}
\bibliography{paper.bib}

\let\textttOrig=\texttt
\def\texttt#1{\expandafter\textttOrig{\seqsplit{#1}}}
\renewcommand{\seqinsert}{\ifmmode
  \allowbreak
  \else\penalty6000\hspace{0pt plus 0.02em}\fi}

\makeatletter
\let\href@Orig=\href
\def\href@Urllike#1#2{\href@Orig{#1}{\begingroup
    \def\Url@String{#2}\Url@FormatString
    \endgroup}}
\def\href@Notdoi#1#2{\def\tempa{#1}\def\tempb{#2}%
  \ifx\tempa\tempb\relax\href@Urllike{#1}{#2}\else
  \href@Orig{#1}{#2}\fi}
\def\href#1#2{%
  \IfBeginWith{#1}{https://doi.org}%
  {\href@Urllike{#1}{#2}}{\href@Notdoi{#1}{#2}}}
\makeatother

\newlength{\cslhangindent}
\newlength{\csllabelwidth}
\newenvironment{CSLReferences}[3] 
 {
  \setlength{\parindent}{0pt}
  \ifodd #1 \everypar{\setlength{\hangindent}{\cslhangindent}}\ignorespaces\fi
  \ifnum #2 > 0
  \setlength{\parskip}{#2\baselineskip}
  \fi
 }%
 {}
\usepackage{calc}

\usepackage[top=3.5cm, bottom=3cm, right=1.5cm, left=1.0cm,
            headheight=2.2cm, reversemp, includemp, marginparwidth=4.5cm]{geometry}

\titleformat{\section}
  {\normalfont\sffamily\Large\bfseries}
  {}{0pt}{}
\titleformat{\subsection}
  {\normalfont\sffamily\large\bfseries}
  {}{0pt}{}
\titleformat{\subsubsection}
  {\normalfont\sffamily\bfseries}
  {}{0pt}{}
\titleformat*{\paragraph}
  {\sffamily\normalsize}

\usepackage{fancyhdr}
\makeatletter
\let\ps@plain\ps@fancy
\fancyheadoffset[L]{4.5cm}
\fancyfootoffset[L]{4.5cm}

\definecolor{linky}{rgb}{0.0, 0.5, 1.0}

\newtcolorbox{repobox}
   {colback=red, colframe=red!75!black,
     boxrule=0.5pt, arc=2pt, left=6pt, right=6pt, top=3pt, bottom=3pt}

\newcommand{\ExternalLink}{%
   \tikz[x=1.2ex, y=1.2ex, baseline=-0.05ex]{%
       \begin{scope}[x=1ex, y=1ex]
           \clip (-0.1,-0.1)
               --++ (-0, 1.2)
               --++ (0.6, 0)
               --++ (0, -0.6)
               --++ (0.6, 0)
               --++ (0, -1);
           \path[draw,
               line width = 0.5,
               rounded corners=0.5]
               (0,0) rectangle (1,1);
       \end{scope}
       \path[draw, line width = 0.5] (0.5, 0.5)
           -- (1, 1);
       \path[draw, line width = 0.5] (0.6, 1)
           -- (1, 1) -- (1, 0.6);
       }
   }

\patchcmd{\@maketitle}{center}{flushleft}{}{}
\patchcmd{\@maketitle}{center}{flushleft}{}{}
\patchcmd{\@maketitle}{\LARGE}{\LARGE\sffamily}{}{}
\def\maketitle{{%
  
  \AB@maketitle}}
\makeatletter
\renewcommand\AB@affilsepx{ \protect\Affilfont}
\renewcommand\AB@affilnote[1]{{\bfseries #1}\hspace{3pt}}
\renewcommand{\affil}[2][]%
   {\newaffiltrue\let\AB@blk@and\AB@pand
      \if\relax#1\relax\def\AB@note{\AB@thenote}\else\def\AB@note{#1}%
        \setcounter{Maxaffil}{0}\fi
        \begingroup
        \let\href=\href@Orig
        \let\texttt=\textttOrig
        \let\protect\@unexpandable@protect
        \def\thanks{\protect\thanks}\def\footnote{\protect\footnote}%
        \@temptokena=\expandafter{\AB@authors}%
        {\def\\{\protect\\\protect\Affilfont}\xdef\AB@temp{#2}}%
         \xdef\AB@authors{\the\@temptokena\AB@las\AB@au@str
         \protect\\[\affilsep]\protect\Affilfont\AB@temp}%
         \gdef\AB@las{}\gdef\AB@au@str{}%
        {\def\\{, \ignorespaces}\xdef\AB@temp{#2}}%
        \@temptokena=\expandafter{\AB@affillist}%
        \xdef\AB@affillist{\the\@temptokena \AB@affilsep
          \AB@affilnote{\AB@note}\protect\Affilfont\AB@temp}%
      \endgroup
       \let\AB@affilsep\AB@affilsepx
}
\makeatother

\renewcommand\Affilfont{\sffamily\small\mdseries}
\ifnum 0\ifxetex 1\fi\ifluatex 1\fi=0 
  \usepackage[T1]{fontenc}
  \usepackage[utf8]{inputenc}

\else 
  \ifxetex
    \usepackage{mathspec}
    \usepackage{fontspec}

  \else
    \usepackage{fontspec}
  \fi
  \defaultfontfeatures{Ligatures=TeX,Scale=MatchLowercase}

\fi
\IfFileExists{upquote.sty}{\usepackage{upquote}}{}
\IfFileExists{microtype.sty}{%
\usepackage{microtype}
\UseMicrotypeSet[protrusion]{basicmath} 
}{}

\usepackage{hyperref}
\hypersetup{unicode=true,
            pdftitle={MCALF: Multi-Component Atmospheric Line Fitting},
            pdfborder={0 0 0},
            breaklinks=true}
\let\addcontentslineOrig=\addcontentsline
\def\addcontentsline#1#2#3{\bgroup
  \let\texttt=\textttOrig\addcontentslineOrig{#1}{#2}{#3}\egroup}
\let\markbothOrig\markboth
\def\markboth#1#2{\bgroup
  \let\texttt=\textttOrig\markbothOrig{#1}{#2}\egroup}
\let\markrightOrig\markright
\def\markright#1{\bgroup
  \let\texttt=\textttOrig\markrightOrig{#1}\egroup}

\usepackage{graphicx,grffile}
\makeatletter
\def\maxwidth{\ifdim\Gin@nat@width>\linewidth\linewidth\else\Gin@nat@width\fi}
\def\maxheight{\ifdim\Gin@nat@height>\textheight\textheight\else\Gin@nat@height\fi}
\makeatother
\setkeys{Gin}{width=\maxwidth,height=\maxheight,keepaspectratio}
\IfFileExists{parskip.sty}{%
\usepackage{parskip}
}{
\setlength{\parindent}{0pt}
\setlength{\parskip}{6pt plus 2pt minus 1pt}
}
\ifx\paragraph\undefined\else
\let\oldparagraph\paragraph
\renewcommand{\paragraph}[1]{\oldparagraph{#1}\mbox{}}
\fi
\ifx\subparagraph\undefined\else
\let\oldsubparagraph\subparagraph
\renewcommand{\subparagraph}[1]{\oldsubparagraph{#1}\mbox{}}
\fi

\title{MCALF: Multi-Component Atmospheric Line Fitting}

        \author[1]{Conor D. MacBride}
          \author[1, 2]{David B. Jess}
    
      \affil[1]{Astrophysics Research Centre, School of Mathematics and
Physics, Queen's University Belfast, Belfast, BT7 1NN, UK}
      \affil[2]{Department of Physics and Astronomy, California State
University Northridge, Northridge, CA 91330, U.S.A.}
  \date{\vspace{-7ex}}

\begin{document}
\maketitle

\marginpar{

  \begin{flushleft}
  \sffamily\small

  {\bfseries DOI:} \href{https://doi.org/10.21105/joss.03265}{\color{linky}{10.21105/joss.03265}}

  \vspace{2mm}

  {\bfseries Software}
  \begin{itemize}
    \setlength\itemsep{0em}
    \item \href{https://github.com/openjournals/joss-reviews/issues/3265}{\color{linky}{Review}} \ExternalLink
    \item \href{https://github.com/ConorMacBride/mcalf}{\color{linky}{Repository}} \ExternalLink
    \item \href{https://doi.org/10.5281/zenodo.4767888}{\color{linky}{Archive}} \ExternalLink
  \end{itemize}

  \vspace{2mm}

  \par\noindent\hrulefill\par

  \vspace{2mm}

  {\bfseries Editor:} \href{http://stanford.edu/~mbobra/}{Monica Bobra} \ExternalLink \\
  \vspace{1mm}
    {\bfseries Reviewers:}
  \begin{itemize}
  \setlength\itemsep{0em}
    \item \href{https://github.com/mgalloy}{@mgalloy}
    \item \href{https://github.com/Goobley}{@Goobley}
    \end{itemize}
    \vspace{2mm}

  {\bfseries Submitted:} 19 April 2021\\
  {\bfseries Published:} 17 May 2021

  \vspace{2mm}
  {\bfseries License}\\
  Authors of papers retain copyright and release the work under a Creative Commons Attribution 4.0 International License (\href{http://creativecommons.org/licenses/by/4.0/}{\color{linky}{CC BY 4.0}}).

  \end{flushleft}
}

\hypertarget{summary}{%
\section{Summary}\label{summary}}

Determining accurate velocity measurements from observations of the Sun
is of vital importance to solar physicists who are studying the wave
dynamics in the solar atmosphere. Weak chromospheric absorption lines,
due to dynamic events in the solar atmosphere, often consist of multiple
spectral components. Isolating these components allows for the velocity
field of the dynamic and quiescent regimes to be studied independently.
However, isolating such components is particularly challenging due to
the wide variety of spectral shapes present in the same dataset.
\texttt{MCALF} provides a novel method and infrastructure to determine
Doppler velocities in a large dataset. Each spectrum is fitted with a
model adapted to its specific spectral shape.

\hypertarget{statement-of-need}{%
\section{Statement of need}\label{statement-of-need}}

MCALF is an open-source Python package for accurately constraining
velocity information from spectral imaging observations using machine
learning techniques. This software package is intended to be used by
solar physicists trying to extract line-of-sight (LOS) Doppler velocity
information from spectral imaging observations (Stokes \(I\)
measurements) of the Sun. This `toolkit' can be used to define a
spectral model optimised for a particular dataset.

This package is particularly suited for extracting velocity information
from spectral imaging observations where the individual spectra can
contain multiple spectral components. Such multiple components are
typically present when active solar phenomena occur within an isolated
region of the solar disk. Spectra within such a region will often have a
large emission component superimposed on top of the underlying
absorption spectral profile from the quiescent solar atmosphere (Felipe
et al., 2014). Being able to extract velocity information from such
observations would provide solar physicists with a wider range of data
products that can be used for science (Stangalini et al., 2020). This
package implements the novel approach of automated classification of
spectral profiles prior to fitting a model.

A sample model is provided for an IBIS Ca \(\text{\sc{ii}}\) 8542 Å
spectral imaging sunspot dataset. This dataset typically contains
spectra with multiple atmospheric components and this package supports
the isolation of the individual components such that velocity
information can be constrained for each component. The method
implemented in this IBIS model has been discussed extensively in
MacBride et al. (2020). There are also several ongoing research projects
using this model to extract velocity measurements.

Using this sample model, as well as the separate base (template) model
it is built upon, a custom model can easily be built for a specific
dataset. The custom model can be designed to take into account the
spectral shape of each particular spectrum in the dataset. By training a
neural network classifier using a sample of spectra from the dataset
labelled with their spectral shapes, the spectral shape of any spectrum
in the dataset can be found. The fitting algorithm can then be adjusted
for each spectrum based on the particular spectral shape the neural
network assigned it. The `toolkit' nature of this package also allows
the possibility of utilising existing machine learning classifiers, such
as the ``supervised hierarchical \(k\)-means" classifier introduced in
Panos et al. (2018), which clusters solar flare spectra based on their
profile shape.

\begin{figure}
\centering
\includegraphics{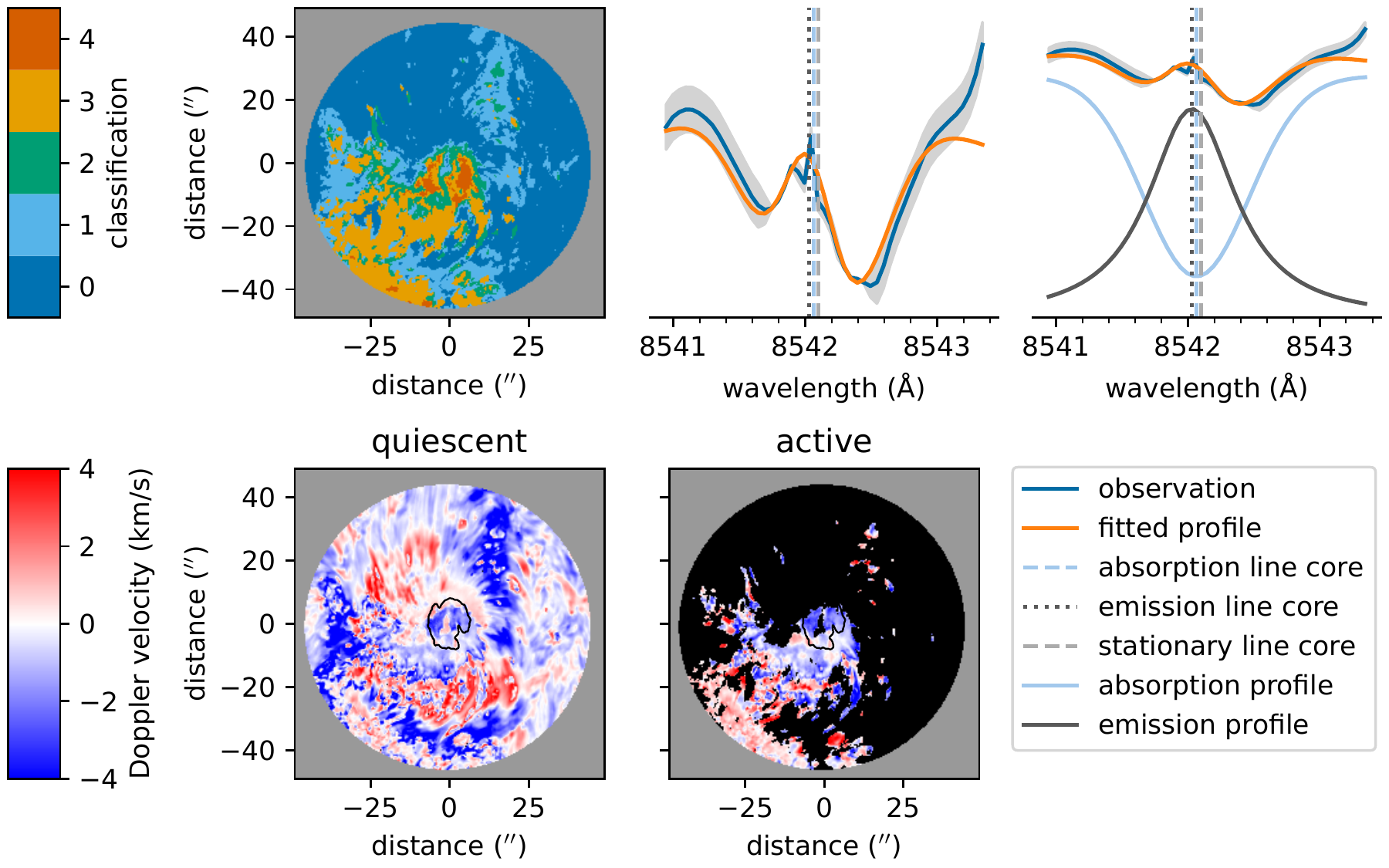}
\caption{An overview of some of the plotting functions that are included
in \texttt{MCALF}.\label{fig:example}}
\end{figure}

This package is designed to run in parallel over large data cubes, as
well as in serial. As each spectrum is processed in isolation, this
package scales very well across many processor cores. Numerous functions
are provided to plot the results clearly, some of which are showcased in
\autoref{fig:example}. The \texttt{MCALF} API also contains many useful
functions which have the potential of being integrated into other Python
packages. Full documentation as well as examples on how to use
\texttt{MCALF} are provided at
\href{https://mcalf.macbride.me}{mcalf.macbride.me}.

\hypertarget{acknowledgements}{%
\section{Acknowledgements}\label{acknowledgements}}

CDM would like to thank the Northern Ireland Department for the Economy
for the award of a PhD studentship. DBJ wishes to thank Invest NI and
Randox Laboratories Ltd.~for the award of a Research and Development
Grant (059RDEN-1) that allowed the computational techniques employed to
be developed. DBJ would also like to thank the UK Science and Technology
Facilities Council (STFC) for the consolidated grant ST/T00021X/1. The
authors wish to acknowledge scientific discussions with the Waves in the
Lower Solar Atmosphere (WaLSA;
\href{https://www.WaLSA.team}{www.WaLSA.team}) team, which is supported
by the Research Council of Norway (project no. 262622) and the Royal
Society (award no. Hooke18b/SCTM).

\hypertarget{references}{%
\section*{References}\label{references}}
\addcontentsline{toc}{section}{References}

\hypertarget{refs}{}
\begin{CSLReferences}{1}{0}
\leavevmode\hypertarget{ref-Felipe:2014}{}%
Felipe, T., Socas-Navarro, H., \& Khomenko, E. (2014). Synthetic
observations of wave propagation in a sunspot umbra. \emph{The
Astrophysical Journal}, \emph{795}(1), 9.
\url{https://doi.org/10.1088/0004-637x/795/1/9}

\leavevmode\hypertarget{ref-MacBride:2020}{}%
MacBride, C. D., Jess, D. B., Grant, S. D. T., Khomenko, E., Keys, P.
H., \& Stangalini, M. (2020). Accurately constraining velocity
information from spectral imaging observations using machine learning
techniques. \emph{Philosophical Transactions of the Royal Society A:
Mathematical, Physical and Engineering Sciences}, \emph{379}(2190),
20200171. \url{https://doi.org/10.1098/rsta.2020.0171}

\leavevmode\hypertarget{ref-Panos:2018}{}%
Panos, B., Kleint, L., Huwyler, C., Krucker, S., Melchior, M., Ullmann,
D., \& Voloshynovskiy, S. (2018). Identifying typical {Mg} {\textsc{ii}}
flare spectra using machine learning. \emph{The Astrophysical Journal},
\emph{861}(1), 62. \url{https://doi.org/10.3847/1538-4357/aac779}

\leavevmode\hypertarget{ref-Stangalini:2020}{}%
Stangalini, M., Baker, D., Valori, G., Jess, D. B., Jafarzadeh, S.,
Murabito, M., To, A. S. H., Brooks, D. H., Ermolli, I., Giorgi, F., \&
MacBride, C. D. (2020). Spectropolarimetric fluctuations in a sunspot
chromosphere. \emph{Philosophical Transactions of the Royal Society A:
Mathematical, Physical and Engineering Sciences}, \emph{379}(2190),
20200216. \url{https://doi.org/10.1098/rsta.2020.0216}

\end{CSLReferences}

\end{document}